\documentclass[preprint]{aastex}
\usepackage{ulem} 
\usepackage{longtable}
\usepackage{rotating}

\shorttitle{Astrometry and Photometry with {\it HST}-WFC3/UVIS. II.}
\shortauthors{Bellini, A. et al.\ }

\begin{document}

\title{Astrometry and photometry with {\it HST} WFC3. II.  Improved
geometric-distortion corrections for 10 filters of the
UVIS channel\footnote{Based on observations with the NASA/ESA {\it
Hubble Space Telescope}, obtained at the Space Telescope Science
Institute, which is operated by AURA, Inc., under NASA contract NAS
5-26555.\newline}}

\author{A. Bellini\footnote{ Visiting Ph.D.\ Student at STScI under
    the {\it ``2008 graduate research assistantship''}
    program.}\affil{Dipartimento di Astronomia, Universit\`a di
    Padova, Vicolo dell'Osservatorio 3, Padova, I-35122, Italy,
    EU}\email{andrea.bellini@unipd.it}} 
\author{J. Anderson, L.~R. Bedin
  \affil{Space Telescope Science Institute, 3700 San Martin Drive,
    Baltimore, MD 21218, USA}\email{jayander,bedin@stsci.edu}}

\begin{abstract}
We present an improved geometric-distortion solution for the
\textit{Hubble Space Telescope} UVIS channel of Wide Field Camera 3
for ten broad-band filters. The solution is made up of three
parts:\ (1) a 3$^{\rm rd}$-order polynomial to deal with the general
optical distortion, (2) a table of residuals that accounts for both
chip-related anomalies and fine-structure introduced by the filter,
and (3) a linear transformation to put the two chips into a convenient
master frame.  The final correction is better than 0.008 pixel ($\sim
0.3$ mas) in each coordinate.  We provide the solution in two
different forms: a \texttt{FORTRAN} subroutine and a set of
\texttt{fits} files, one for each filter/chip/coordinate.
\end{abstract}

\section{Introduction, data set, measurements}
\label{sec1}

For various reasons, such as camera optics, alignment errors, filter
irregularities, non-flat CCDs, and CCD manufacturing defects, the
mapping of the square array of square pixels of a detector onto the
tangent-plane projection of the sky requires a non-linear
transformation.  Positions measured within the pixel grid need to be
corrected for geometric distortion (GD) before they can be accurately
compared with other positions in the same image, or compared with
positions measured in other image.

While almost all scientific programs must make use of the distortion
solution, most are relatively insensitive to it.  So long as each
detector pixel is mapped to within a fraction of a pixel of its true
location in a distortion-corrected reference frame, the resampling
accuracy is not compromised.  For this reason, the formal requirement
for distortion calibration for WFC3 was 0.2 pixel to enable
\texttt{MultiDrizzle} in the calibration pipeline to generate stacked
associations.  Kozhurina-Platais et al.\ (2009;\ ISR 2009-33) and
McClean et al.\ (2010;\ ISR 2009-041) have demonstrated that the
current official solution is accurate to 0.05 pixel (2 mas), clearly
meeting the needs of the pipeline.

Scientific programs that are focused on astrometry make more stringent
demands on the distortion solution.  These programs must analyze the
raw, un-resampled pixels of the distorted images in order to attain
the highest possible position accuracy.  Similarly, these programs
require a more accurate distortion solution to relate measured
positions to one another.  In general, positions can be measured to a
precision of 0.01 pixel for a well-exposed star, so we would like to
have a distortion solution that is at least this accurate.  Such a
precision is well below the formal calibration requirements, but if it
can be shown that the solution is stable at that level, then an
improved calibration will enable many astrometry-related programs.

In a previous paper (Bellini \& Bedin 2009, hereafter Paper~I), we
used the limited set of dithered exposures taken during SMOV
(Servicing Mission Observatory Verification) and an existing ACS/WFC
astrometric reference frame as a flat-field to derive a set of 3$^{\rm
  rd}$-order-polynomial correction coefficients to represent the
geometric distortion in WFC3/UVIS.  The solution was derived
independently for each of the two CCDs for each of the three
broad-band ultraviolet filters F225W, F275W and F336W.

We found that by applying our correction it was possible to remove the
GD over the entire area of each chip to an average 1-D accuracy of
0.025 pixels (i.e., 1 mas).  At the time, the lack of sufficient
observations collected at different roll angles and dithers prevented
a more accurate self-calibration-based solution.

The calibration data collected over the past year has enabled us to
undertake the next step in modeling the GD of WFC3/UVIS detectors.
The large number of dithers and multiple roll angles now available
allow us to construct a 2010-based reference frame that is free of the
proper-motion (PM) errors inherent in the previous 2002-epoch ACS
catalog\footnote{ There were PMs available for some of the stars in
  Anderson \& van der Marel (2010), but the field coverage was not
  complete.}.  The calibration data has enabled us to extend the GD
solution to the rest of the major broad-band filters and to improve
the accuracy of the solution to about 0.008 pixel.  This is below the
precision with which we can measure a well-exposed stars in a single
exposure.  Improvements beyond this are unlikely, as breathing and
other temporal variations make a more accurate ``average'' solution
unnecessary.

As we will see in Sect.~\ref{sec2}, the UVIS chips contain distortion
with very complex variations on relatively small spatial scales, which
would be very unwieldy to model with simple polynomials.  Therefore we
chose to model the GD with two components: a simple 3$^{\rm rd}$-order
polynomial and an empirically-derived table of residuals (the look-up
table).

These look-up tables (one per chip/filter combination) simultaneously
absorb {\it i) } the complicated effects caused by the non-perfectly
uniform and flat surfaces of the filters (for an example in the case
of ACS/WFC, see Anderson \& King 2006, hereafter AK06) and {\it ii)}
the astrometric imprint of manufacturing defects on the UVIS CCD pixel
arrays.  This approach was pursued with success on both ACS/HRC and
ACS/WFC detectors (Anderson 2002, AK06), for which the GD correction
based on polynomials plus look-up tables is still the best available
to date.

The data used in this paper come from the $\omega$~Cen calibration
field, and were collected under programs PID-11452 (PI-Quijano) and
PID-11911 (PI-Sabbi) during 4 epochs:\ 2009 July 15 (2009.54), 2010
January 12--14 (2010.04), 2010 April 28--29 (2010.33) and 2010 June
30--July 4 (2010.51).  During the January and April 2010 runs
guide-star failures made it necessary to repeat some of the exposures
of PID-11911. Moreover, between the first and the second epoch the
telescope focus changed significantly.  The summary of the
observations used in this work is given in Table~\ref{tab1}.  In
addition to large dithers of the size of $\sim$1000 pixels
(i.e.\ $\sim$40$^{\prime\prime}$), at least 2 orientations were
available for each filter.

Star positions and fluxes were obtained as described in Paper~I, using
a spatially-variable PSF-fitting method, which will be described in
detail in a subsequent paper of this series.

\begin{center}
\begin{table}[th!]
\caption{Log of $\omega$~Cen observations used in this work, sorted by
  epoch (PID-11452, PID-11911).}
\label{tab1}
\footnotesize{
\begin{tabular}{ccccccc}
\hline \hline
\textbf{Epoch}&\textbf{PA\_V3}&\textbf{F225W} & \textbf{F275W} & \textbf{F336W}  & \textbf{F390W}  & \textbf{F438W}\\
\hline
 2009.54       & $286^\circ$ &                 & $3\times350\,$s & $8\times350\,$s &                 & \\
 2010.04       & $105^\circ$ & $9\times900\,$s & $9\times800\,$s & $5\times350\,$s & $8\times350\,$s & $8\times350\,$s \\
 2010.33       & $200^\circ$ & $5\times900\,$s &                 & $1\times350\,$s & $5\times350\,$s & $7\times350\,$s \\
 2010.51       & $280^\circ$ &                 & $9\times800\,$s & $9\times350\,$s &                 & $9\times350\,$s \\
\hline
\textbf{Epoch}&\textbf{PA\_V3}&\textbf{F555W} & \textbf{F606W} & \textbf{F775W}  & \textbf{F814W}  & \textbf{F850LP}\\
\hline
 2009.54       & $286^\circ$ &                 &                 &                 &                 & \\
 2010.04       & $105^\circ$ &  $6\times40\,$s &  $9\times40\,$s & $9\times350\,$s &  $8\times40\,$s & $9\times60\,$s \\
2010.32 & $200^\circ$ &               & $9\times40\,$s$^{(\ast)}$&          &                 & \\ 
 2010.33       & $200^\circ$ &  $9\times40\,$s &  $4\times40\,$s & $7\times350\,$s &  $6\times40\,$s & $5\times60\,$s \\
 2010.51       & $280^\circ$ &                 &  $9\times40\,$s &                 &  $9\times40\,$s & \\
\hline \hline
\multicolumn{7}{l}{Notes: $^{(\ast)}$ used only to derive the meta-coordinate-frame solution (see Sect.~\ref{sec:global}).}\\
\end{tabular}}
\end{table}
\begin{center}
\end{center}
\begin{table*}[th!]
\caption{Example of coefficients of the 3$^{\rm rd}$-order polynomial
  solution for chip[1] and chip[2] in the case of the F606W filter
  (coefficients for all the filters are listed in Table~A in the
  electronic version of the paper).  Note that the polynomials are
  normalized such that the size of the coefficient corresponds to the
  maximum contribution (in pixels) of that term across the chip.}
\label{tab2}
\footnotesize{
\begin{tabular}{ccrrrr}
\hline
\hline
Term $\!(w)\!\!\!\!\!$&Polyn.$\!\!\!\!\!\!\!\!\!$
&$a_{w,[1]}$&$b_{w,[1]}$&$a_{w,[2]}$&$b_{w,[2]}$\\ 
\hline
1&$\tilde{x}$&            $  0.000 \!$&$\! 129.384 \!$&$\!  0.000  \!$&$\! 140.280 \!$\\
2&$\tilde{y}$&            $  0.000 \!$&$\!   1.868 \!$&$\!  0.000  \!$&$\!  -4.251 \!$\\
3&$\tilde{x}^2$&          $ 12.044 \!$&$\!   0.622 \!$&$\! 11.960  \!$&$\!   0.762 \!$\\
4&$\tilde{x}\tilde{y}$&   $ -6.222 \!$&$\!   5.526 \!$&$\! -6.069  \!$&$\!   5.462 \!$\\
5&$\tilde{y}^2$&          $  0.065 \!$&$\!  -3.211 \!$&$\!  0.001  \!$&$\!  -3.063 \!$\\
6&$\tilde{x}^3$&          $  0.183 \!$&$\!   0.026 \!$&$\!  0.096  \!$&$\!   0.156 \!$\\
7&$\tilde{x}^2\tilde{y}$& $ -0.054 \!$&$\!   0.080 \!$&$\!  0.037  \!$&$\!  -0.009 \!$\\
8&$\tilde{x}\tilde{y}^2$& $  0.033 \!$&$\!  -0.034 \!$&$\!  0.076  \!$&$\!  -0.020 \!$\\
9&$\tilde{y}^3$&          $  0.023 \!$&$\!  -0.009 \!$&$\!  0.036  \!$&$\!   0.049 \!$\\
\hline
\hline
\end{tabular}}
\end{table*}
\end{center}

\begin{figure*}[t!]
\centering
\includegraphics[width=14cm]{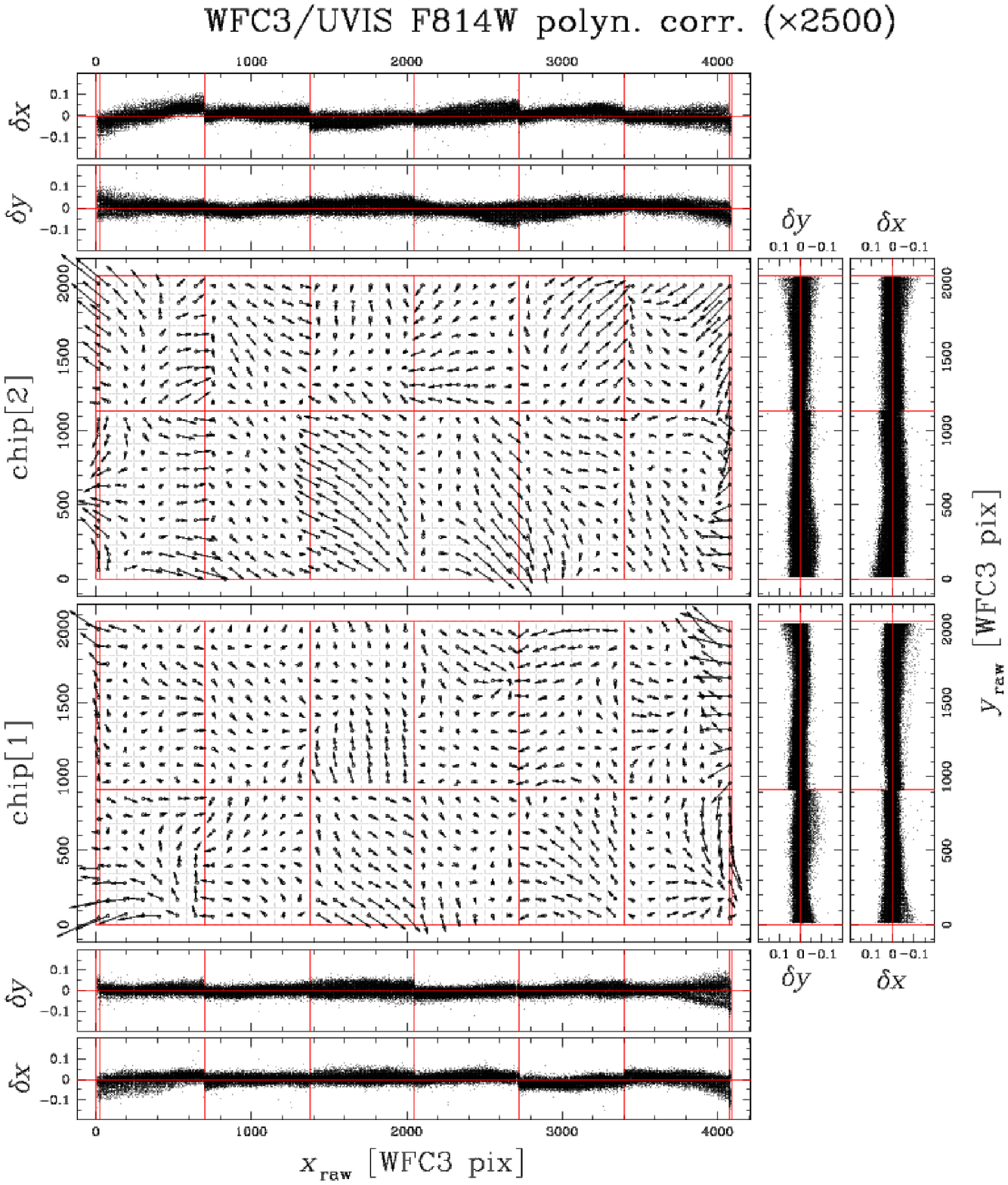}
\caption{Positional residuals ---in the raw-pixel-coordinate system---
  between the master frame, obtained modeling the GD with {\it only}
  our 3$^{\rm rd}$-order polynomial, and the raw position measured on
  individual exposures.  These residuals are averaged within each cell
  and displayed as arrow vectors (magnified by a factor of 2500).
  This figure reveals clear discontinuities that cannot be modeled by
  polynomials of any reasonable order.  These discontinuities are
  marked with red solid lines, which are used to define edges on the
  look-up table (see the text for more details).}
\label{fig:res_before}
\end{figure*}

\section{The back-bone third-order polynomial solution}
\label{sec2}

The GD solution was constructed in three stages.  We began by treating
each chip independently.  We first solved for the 3$^{\rm rd}$-order
polynomial that provides the lion's share of the correction for each
chip.  After subtracting this global component of the solution, we
were then able to see and model the fine-scale component of the
solution, caused by detector and filter irregularities.  Finally, with
the solution for each chip nailed down, we found the global linear
parameters that mapped both chips into a convenient meta-chip
reference frame.

In Paper~I we derived a 3$^{\rm rd}$-order polynomial solution for the
GD for filters F225W, F275W and F336W; here our aim in this section is
to obtain the polynomial solution for seven other broad-band filters
for which suitable observations are available, namely: F390W, F438W,
F555W, F606W, F775W, F814W and F850LP.

Whereas in Paper~I we had to use an astrometric 
reference catalog that was taken several years prior in order to
extract the GD solution, we can now perform a self-calibration of the
GD thanks to the improved number of images at different roll angles
offered by the new data set.  Self-calibrations can often be more
accurate than calibrations that reference a standard field, since
stars in standard fields can move due to proper motions and since the
brightness range where stars in the catalog are well measured may not
correspond to the brightness range where stars in the calibration
exposures are well measured.  Furthermore, the images may have
different crowding issues and measurement qualities from the catalog.

We followed the prescriptions given in Anderson \& King (2003) for
WFPC2 and subsequently used by the same authors to derive the GD
correction for the ACS/HRC (Anderson \& King 2004) and for the ACS/WFC
(AK06).  The same strategy was also used by two of us to calibrate a
ground-based instrument (see Bellini \& Bedin 2010 for details).

Briefly, we started with the F336W GD-solution of Paper~I as first
guess to correct star positions for the seven redward of F336W (from
F390W to F850LP) and created a master frame for each filter
independently.  We then performed the iterative procedure described in
Paper~I to improve the polynomial coefficients.  With a better GD
solution, we then re-constructed the master frames and repeated the
entire process three times.  A fourth repetition of the procedure
provided negligible improvement.  The 3$^{\rm rd}$-order polynomial
coefficients for all 10 of the broad-band filters are hard-coded in
the \texttt{FORTRAN} subroutine available at the website described
given in Section~\ref{sec:conc}.  As an example, the coefficients for
filter F606W are reported in Table~\ref{tab2} (coefficients for all
the ten filters are shown in Table~A in the electronic version of the
paper).

\section{The fine-scale geometric-distortion table}
\label{sec3}

The next step in the procedure was to examine the residuals from the
polynomial solution.  To do this we transformed the master-frame
position for each star back into the raw frame of each exposure using
a conformal linear transformation\footnote{This has four parameters:
  an x-offset, a y-offset, a rotation and a scale.}  and the inverse
distortion solution.  The residual was then the difference between
where the star was observed in the frame $(x_{\rm raw},y_{\rm raw})$
and where it should have been in that frame, according to the master
frame.  In Figure~\ref{fig:res_before} we plot this residual as a
function of raw chip coordinate.  Each black dot represents a single
star measured in single exposure.  We combine all the exposures
together to see the overall trends.

Individual residuals were then averaged within small ($\sim$113-pixel
size) cells (conveniently defined as described in the next section)
and are plotted as arrow vectors (magnified by a factor of 2500).
These residuals exhibit a pattern with unexpected abrupt
discontinuities (denoted with the solid red lines).  We note that
while some of the large-scale trends could be partially removed by
adopting higher-order polynomials, these discontinuities, which have
about the same amplitude as the remaining large-scale trends
($\sim$0.02 pixels), would still remain.  These residuals are very
similar to those seen in Kozhurina-Platais et al.\ (2010).

It is interesting to note that these $\sim$0.02 pixel systematic
trends are perfectly consistent with the larger-than-expected residual
dispersion already noted in Paper~I.  These fine-scale trends were
simply washed out by the large internal motions of the $\omega$~Cen
stars ($\sim$0.15 WFC3/UVIS pixel) over the 7-year baseline between
the reference-frame observations of GO-9442 in 2002 and the WFC3/UVIS
observations of PID-11452 in 2009.
 
\begin{figure*}[t!]
\centering \includegraphics[width=14cm]{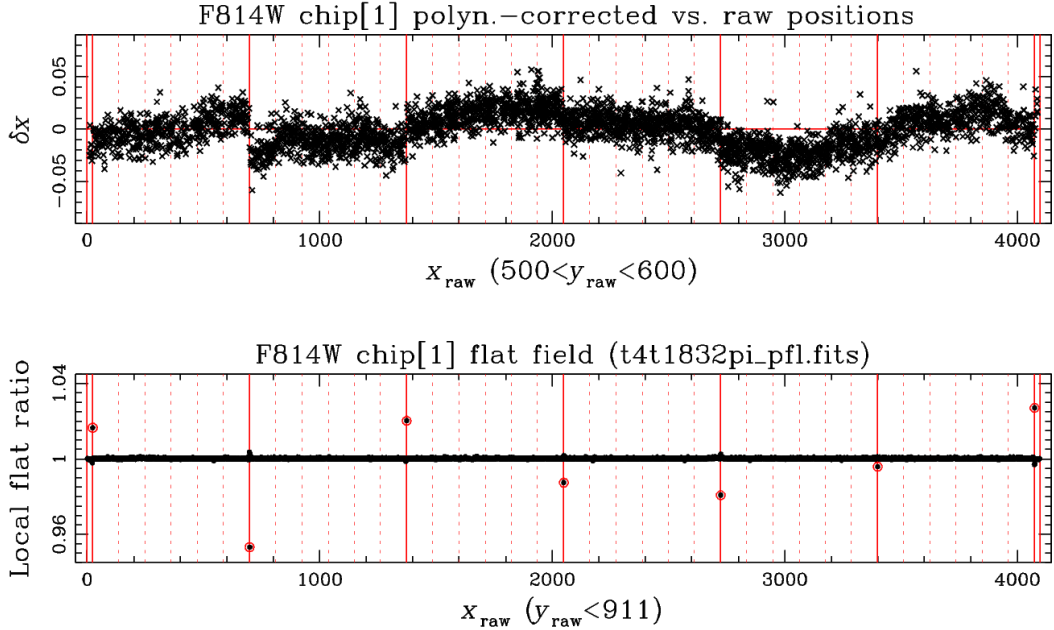}
\caption{
\textit{Top:} Residuals ($\delta x$) between polynomial-corrected and
              raw positions plotted against
              $x_{\rm raw}$ for stars with $500<y_{\rm raw}<600$, in
              chip[1] for the F814W exposures.
              Solid lines mark lithographic-pattern discontinuities.
              Dashed lines show the sampling of our look-up table.
              The 68.27$^{\rm th}$ percentile of $\delta x$ for these
              stars is $\sim$0.02 pixel.  It is clear from this figure
              that uncorrected high-order polynomial residuals and
              lithographic-pattern residuals have about the same
              amplitude.
\textit{Bottom:} 
              For each pixel in chip[1] flat field, we took 
              the ratio of the pixel value over the median of the 30
              pixel values on either side along $x_{\rm raw}$
              directions (independently for each of the two
              amplifiers).  We then took a median of this ratio for
              all the $y_{\rm raw}<911$ pixels in each column. This is
              plotted for each of the 4096 columns.  
              It is clear that there are large local
                    flat-field ratio variations at the locations 
                    of the lithographic-pattern discontinuities.  
                    (highlighted by red open circles),
                    up to $\sim4$\%.}
   \label{fig:fetta}
\end{figure*}

The top panel of Fig.~\ref{fig:fetta} shows the residuals
$\delta x$ 
plotted against
the $x_{\rm raw}$ coordinates, for stars measured within a 
100-pixel-tall strip, centered at $y_{\rm raw}=550$, extracted 
from chip[1]\footnote{See Fig.~\ref{fig:res_before}.  
                            This is the bottom chip, and the first 
                            extension in the \texttt{\_flt} file.
                            This chip is named UVIS2, but to avoid 
                            ambiguity (and to maintain the convention
                            in Paper~I) we will refer to it with 
                            brackets, according to its order within 
                            the \texttt{fits}-image extensions.}.

Constraining the $\delta x$ residuals within a vertical strip
highlights the sharp changes in the residual trend.  It is important
to note that these discontinuities are present in \textit{both} axes
of the WFC3/UVIS detectors and not only along a single axis, as was
the case for WFPC2 (Anderson \& King 1999) or ACS/WFC (Anderson 2002).

The bottom panel of Fig.~\ref{fig:fetta} shows that these boundaries
show up as single-row excursions of up to $\sim4$\% in the F814W flat
fields.  This was also seen for WFPC2 and ACS/WFC.

These discontinuities can be explained as small manufacturing defects
in the CCDs, analogous to those found for the WFPC2 and ACS/WFC
detectors.  These defects arose from an imperfect alignment between
the silicon wafer and the mask used to generate the CCDs' pixel
boundaries during lithographic projection (see Kozhurina-Platais et
al.\ 2010).  Indeed, at the location where these repositioning errors
are found, pixels are wider or narrower in one direction.  As a
consequence, these pixels are respectively brighter or fainter on the
flat field, since they collect more or less light.  This also leads to
the observed astrometric discontinuities.

The $y_{\rm raw}$-axis pattern of these lithographic features on
WFC3/UVIS CCDs are a single pixel wide in the flat field and have a
period of 675 pixels along the $x_{\rm raw}$ axis for both chips,
extending back and forth from the central position $x_{\rm raw} =
2048.5$.\footnote{Note that in our notation we give half values to the
  center of pixels such as, for instance, the center of pixel (1,1) is
  located at $x=0.5$, $y=0.5$ pixels.}  Note that this implies that we
are left with the first and last 23.5-pixel-wide vertical strips where
the discontinuity is repeated.  The horizontal feature along the
$y_{\rm raw}$ axis consists of a single 2-pixel-wide discontinuity,
centered at $y_{\rm raw} = 911$ pixels for chip[1] and at $y_{\rm raw}
= 1140$ pixels for chip[2].  These discontinuities are located
symmetrically with respect to the gap between the two chips
(horizontal solid red lines in Fig.~\ref{fig:res_before}).  A careful
look at Fig.~\ref{fig:fetta} reveals hints of finer discontinuities on
both astrometry (top panel) and flat field (bottom panel), but at much
lower amplitudes (a few thousandths of a pixel and $\sim0.1$\%),
however it's hard to assess their significance.

Our aim here was to find a simple correction of the GD following the
basic principle: ``we see a systematic error, and we empirically find
a correction for it''.  We therefore decided to keep a polynomial of
the third order, as we did in Paper~I, to remove most of the GD (down
to the $\sim1$ mas level), and then to use a single look-up table (one
for each chip and filter) to correct all the smaller-scale positional
systematic errors.  This approach was able to provide accuracies down
to 0.008 pixel ($\sim0.3$ mas), as we will show in the next section.

We set up the look-up table as follows. First we defined $x_{\rm raw}$
boundaries alongside the lithographic feature discontinuities (red
solid lines in Fig.~\ref{fig:res_before}).  We subdivided each of the
675-$x_{\rm raw}$-pixel-wide regions into six 112.5-pixel-wide sub-regions,
for a total of 36 such subdivisions plus two 23.5-pixel-wide
regions at the left and right
edges of each chip.  To maintain a similar sampling along the 
$y_{\rm raw}$ axis, we defined 18 sub-divisions.  As the horizontal
component of the lithographic feature divides the two chips in two
parts of 911 and 1140 pixels tall, we made 8 subdivisions for the
911-$y_{\rm raw}$-pixel one (113.875 pixels each) and 10 for the other
(114 pixels each).  At the end we produced an array of $36\times18$
almost-square cells, plus two 23.5-pixel-wide strips at the short
edges of each chip, each made of 18 rectangular cells (grey dashed
lines in Figs.~\ref{fig:res_before}).

Cell dimensions were ultimately dictated by the necessity to have
enough grid-points to finely sample the GD and to have an adequate
number of stars within each cell to robustly measure the value of the
grid points in the look-up table.  We always had more than 30 stars in
each cell to constrain the value of the table, even for the filters
with the fewest number of well-exposed stars.  Typically we had well
over 100 stars per cell.

\begin{figure*}[t!]
\centering \includegraphics[width=14cm]{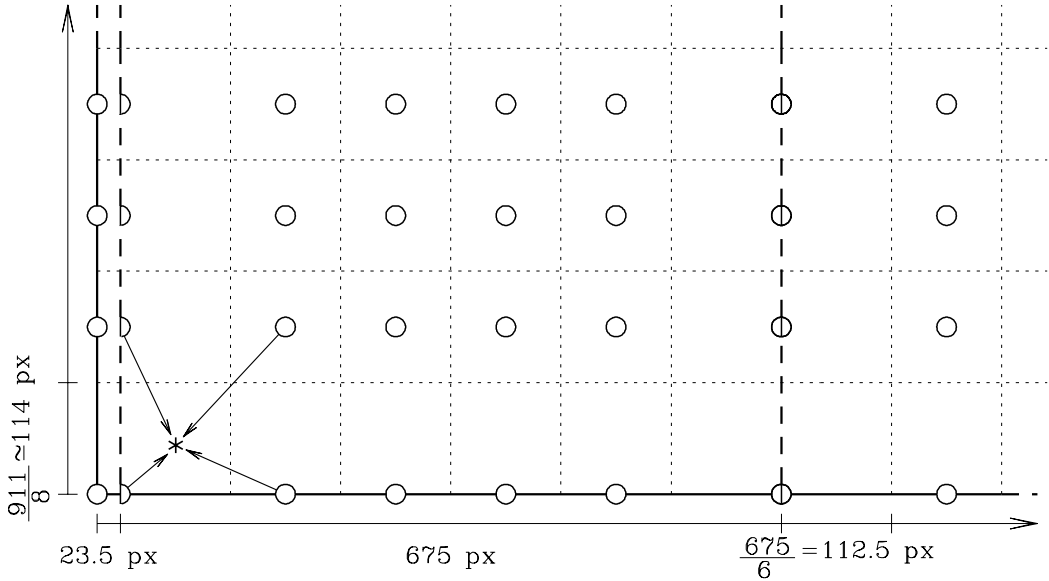}
\caption{Example of cell and grid point locations on the detector.
  This figure shows the bottom-left area of chip[1].  Dotted lines
  denote those regions used to evaluate the amount of the GD residual
  at the grid point locations (empty circles).  Thick solid lines mark
  detector edges and dashed lines identify lithographic
  discontinuities.  The arrows illustrate the process of interpolation
  to evaluate the fine-scale GD residual at a particular location
  (marked with an $\ast$) using the four closest grid-points. Cell
  dimensions are also indicated. (See the text for details).}
\label{fig:scheme}
\end{figure*}

Figure~\ref{fig:scheme} shows an example of the geometry adopted for
the look-up tables.  Thick solid lines mark detector edges, dashed
lines identify lithographic discontinuities, while dotted lines
highlight cell borders.  We used stars within each cell to compute
$3\sigma$-clipped median positional residuals $\overline{\delta x}$
and $\overline{\delta y}$, which are assigned to the corresponding
grid point (open circles).  When a cell adjoins either detector edges
or lithographic discontinuities, the grid point is displaced to the
edge of the cell, as shown.  We use a semicircle to indicate when a
grid point corresponds to a discontinuity.

For any given location on the chip, the look-up table correction is
given by a bi-linear interpolation among the surrounding four grid 
points (as illustrated by the arrows in Fig.~\ref{fig:scheme}).
In the two 23.5-pixel-wide strips the correction is given by a linear
interpolation of the two closest grid points along the $y_{\rm raw}$
axis.

To derive the look-up table we used a master frame which is itself
affected by these lithographic features, since it was constructed from
positions that had not been corrected for them.  Once we had a first
estimate of the tabular corrections, we re-determined an improved
master frame by correcting our raw catalogs with both the polynomial
and the look-up table component.  We repeated the whole process of
building up master frames and improving the table values three times.
A fourth iteration proved to offer negligible improvement.

\begin{figure*}[t!]
\centering
\includegraphics[width=14cm]{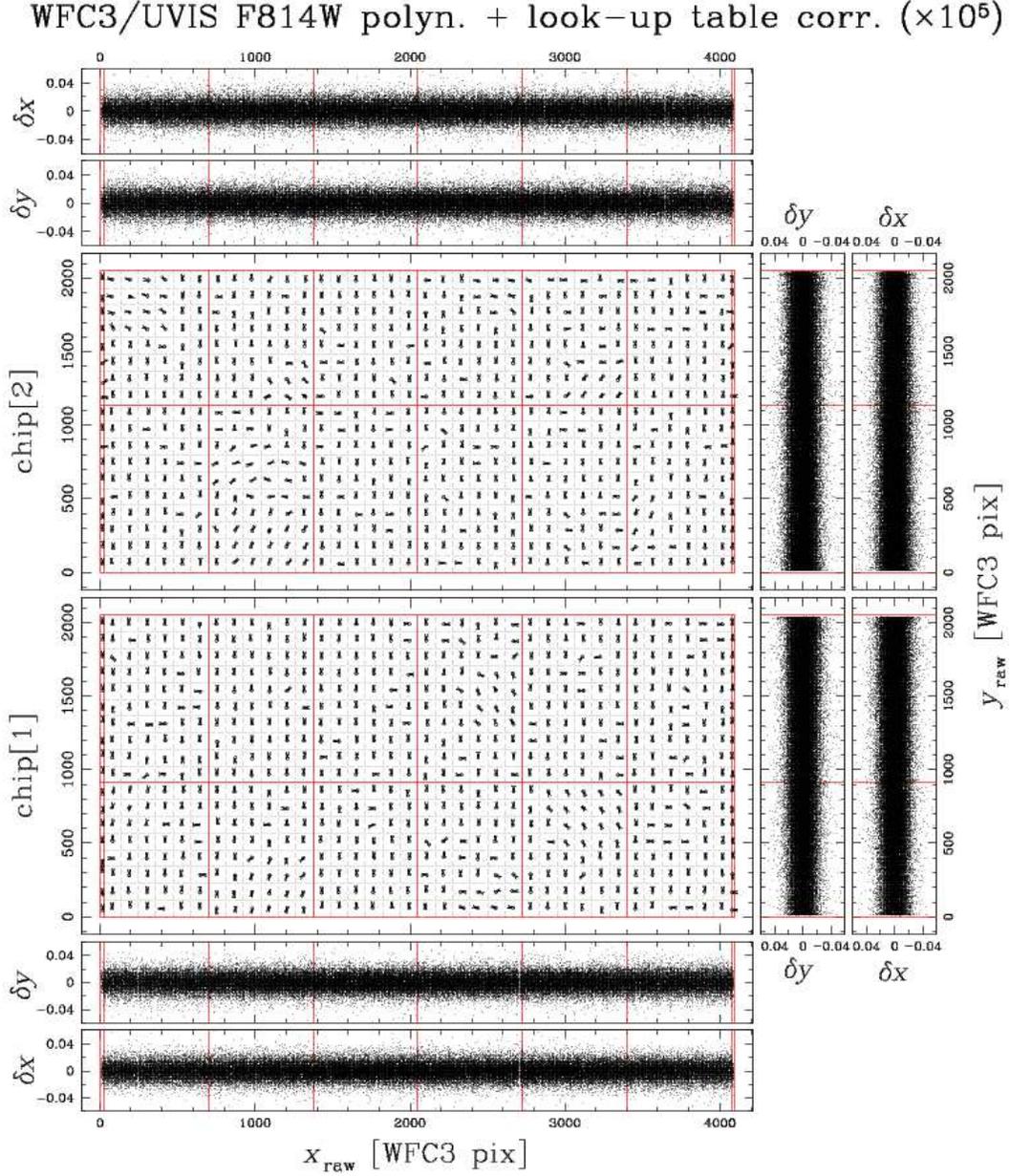}
\caption{Same as Fig.~\ref{fig:res_before}, but after applying the
         look-up table correction.  The size of the residual 
         vectors is now magnified by a factor of $100\,000$.  The 
         lithographic pattern has been completely removed.}
\label{fig:res_after}
\end{figure*}

Figure~\ref{fig:res_after} shows positional residuals in the raw
reference system, after our final look-up table plus polynomial GD
correction is applied.  Residual vectors are now magnified by a factor
100$\,$000.  All the lithographic features and all other
high-frequency patterns seen on Fig.~\ref{fig:res_before} appear to be
completely removed.

\begin{figure*}[t!]
\centering \includegraphics[width=14cm]{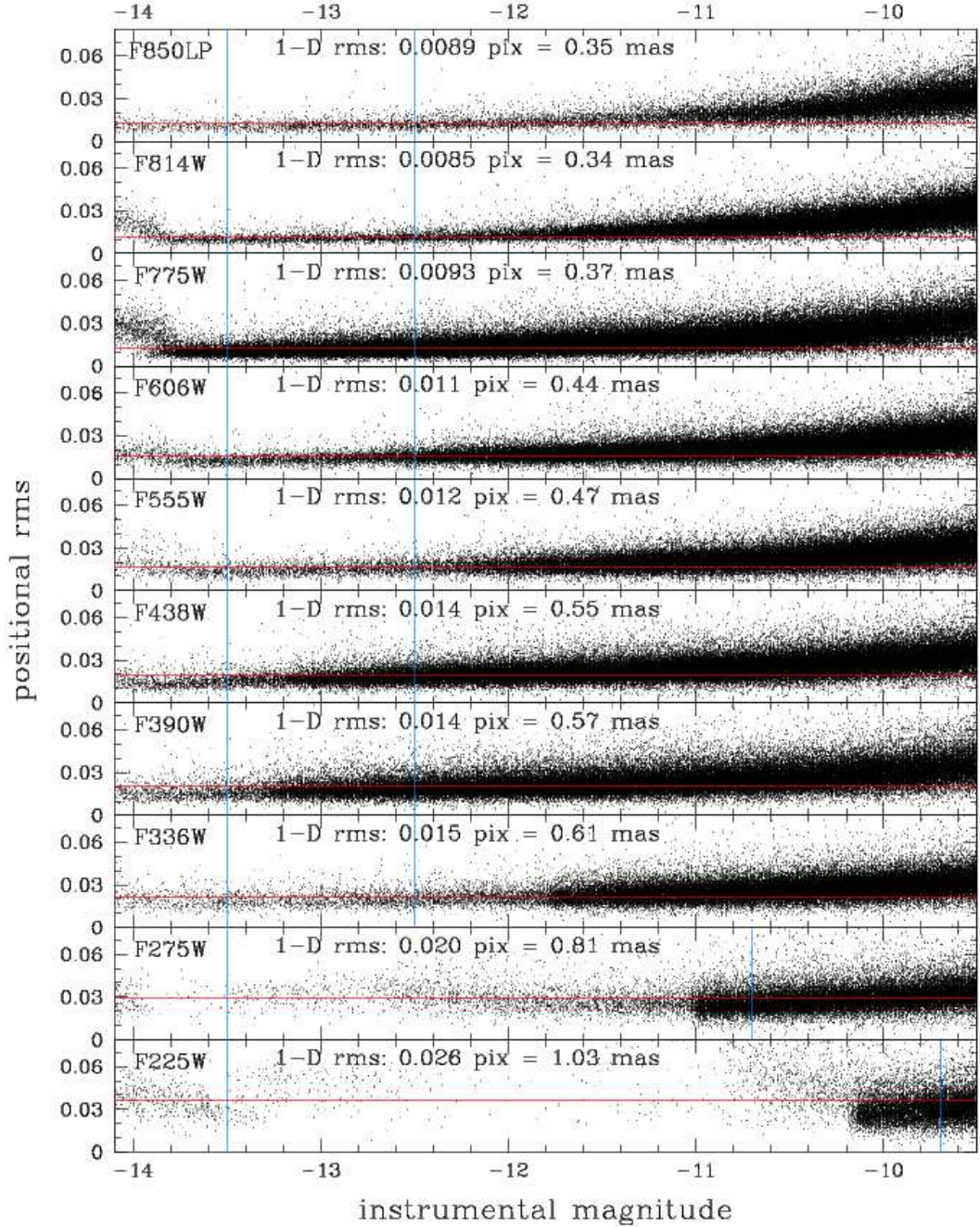}
\caption{RMS of the positional residuals as a function of instrumental
  magnitude for each filter. Saturation typically sets in brightward
  of m$_{\rm instr.}\sim-13.5$.  For well-exposed stars (those within
  the blue boundaries), we computed the 3$\sigma$-clipped 68.27$^{\rm
    th}$ percentile of the positional rms (red lines), for which the
  1-D values are reported on top of each panel in units of both
  WFC3/UVIS pixel and mas.}
\label{fig:res}
\end{figure*}

\begin{figure*}[t!]
\centering
\includegraphics[width=14cm]{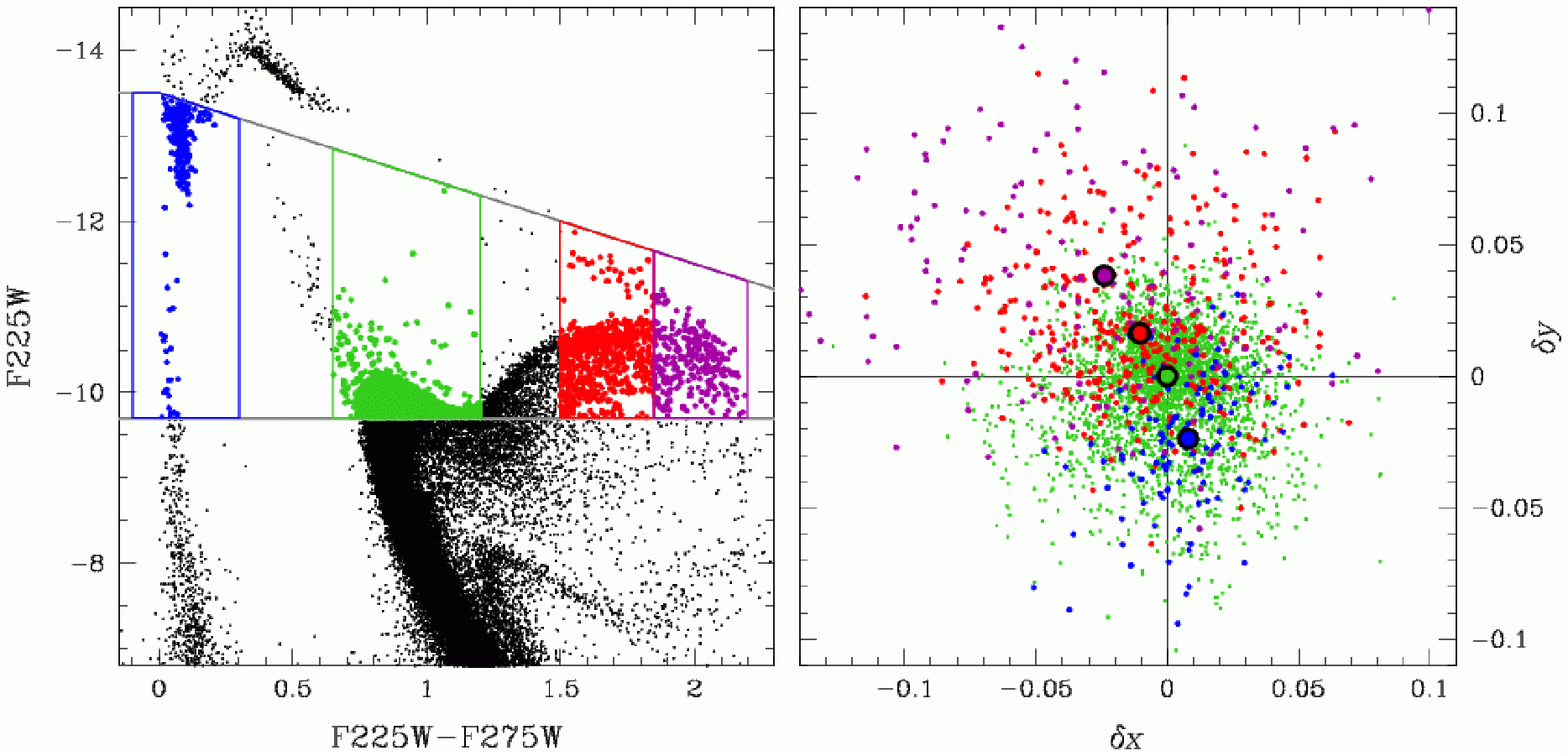}
\caption{\textit{Left:} instrumental F225W vs. F225W$-$F275W CMD. 
                 We selected only images obtained during the same
                 epoch 2010.04 and transformed the F275W master frame
                 positions into F225W one using only
                 unsaturated stars brighter than F225W$=-9.7$ and with
                 F225W$-$F275W color within 0.65 and 1.2 (in green).
                 We isolated 3 other groups of stars (marked in blue,
                 red and purple).
         \textit{Right:} vector-point diagram of the displacements 
                 between the two master frames.  Stars are color
                 coded in the same way as for the left panel. 
                 Full circles mark the median position of each group
                 of stars (median errors have the same size of the 
                 circles). There is a clear systematic trend of the
                 displacements as a function of the stars' color.}
         \label{fig:brtest}
\end{figure*}

One of the best estimates of the true errors of the GD solution is
given by the magnitude of the dispersion (computed as the 
68.27$^{\rm th}$ percentile) of the positional residuals (rms) 
of each star ($i$) observed in each image ($j$), which have been 
GD-corrected and transformed into the master frame 
$(x_{{\rm master},i}; y_{{\rm master},i})$.  These are computed 
as:
$$ 
  \textrm{positional rms$_{i,j}$ [WFC3/UVIS pixels]} = 
          \sqrt{ (x_i^{T_{j}}-x_{{\rm master},i})^2 
               + (y_i^{T_{j}}-y_{{\rm master},i})^2 }.
$$ 
Only stars observed at least in 3 individual images are used to
compute the rms.

Figure~\ref{fig:res} shows these dispersions as a function of the
corresponding instrumental magnitude for different
filters.\footnote{The instrumental magnitude is computed
                            from the sum of the pixel's photo-electrons 
                            under the best fitted PSF 
                            (i.e.\ $-2.5\log{[\Sigma 
                               ({\rm DNs}\times{\rm Gain}})]$).}  

Saturation typically sets in brightward of $\sim$$-$13.5 (marked by
the left blue vertical line). For well-exposed stars (typically
between magnitude $-$13.5 and $-$12.5, but we include fainter stars
for F225W and F275W filters to improve the statistics), the
68.27$^{\rm th}$ percentile levels of the positional rms are marked by
red horizontal lines. The corresponding 1-D values are displayed at
the top of each panel, in units of both pixels and mas.  The best
results are obtained for redder filters.  Here we used all the images
listed in Table~\ref{tab1} (two to four different epochs for each
filter) to compute positional rms.  We will see in the next section
that, by selecting only images within the same epoch, positional
dispersions are even smaller.

We noted that very blue and very red stars behave differently with
respect to our GD correction when observed through the bluest filters
(Kozhurina-Platais et al.\ (2011) see a similar effect).  This is
probably due to a chromatic effect induced by fused-silica CCD windows
within the optical system, which refract differently blue and red
photons and have a sharp increase of the refractive index below 4000
\AA\ (George Hartig, personal communication).  As a consequence, the
F225W, F275W and F336W filters are the most affected.

To better understand how much astrometry will suffer from this
phenomenon, we performed the following test.  We chose the F225W and
F275W data sets for which the effect is maximized and both extreme
horizontal branch (EHB, with a temperature above 40$\,$000$\,$K) and
red giant branch (RGB, with a temperature of $\sim$3000$\,$K) stars
have about the same luminosity (i.e., our positional dispersions have
the same size).  We also required images to be taken within the same
epoch 2010.04 (9 exposures for each filter), to avoid cluster
internal-motion effects.

In the F225W vs. F225W$-$F275W CMD we selected relatively bright
(F225W$<$$-$9.7), unsaturated stars of intermediate color (marked in
green on the left panel of Fig.~\ref{fig:brtest}) and used them to
compute a linear transformation from the F275W master frame into the
F225W frame.  On the same CMD we also selected other groups of stars
with the same luminosity criteria: (i) a blue set made up of EHB and
hot white-dwarf stars; (ii) a red set containing intermediate RGB
stars and (iii) a final purple set populated by RGB-tip stars.
Comparing star positional residuals we found that green stars are
distributed around (0,0), as we would have expected since their
positions formed the basis for the transformations, while stars of the
other groups were found to have residuals located in significantly
different positions (up to $\sim0.04$ pixels away).  On the right
panel of Fig.~\ref{fig:brtest} we show the vector-point diagram for
selected stars, which are color coded as on the left panel of the same
figure.  Median position residuals for each group of stars are marked
by full circles of the same colors.  The size of the circle indicates
the formal error in the median.  A systematic trend of the
displacements as a function of stellar color is clear.  Further
investigations will be required to fully characterize this chromatic
effect.

\section{A demonstrative application}
\label{secdim}

\begin{figure*}[t!]
\centering \includegraphics[width=14cm]{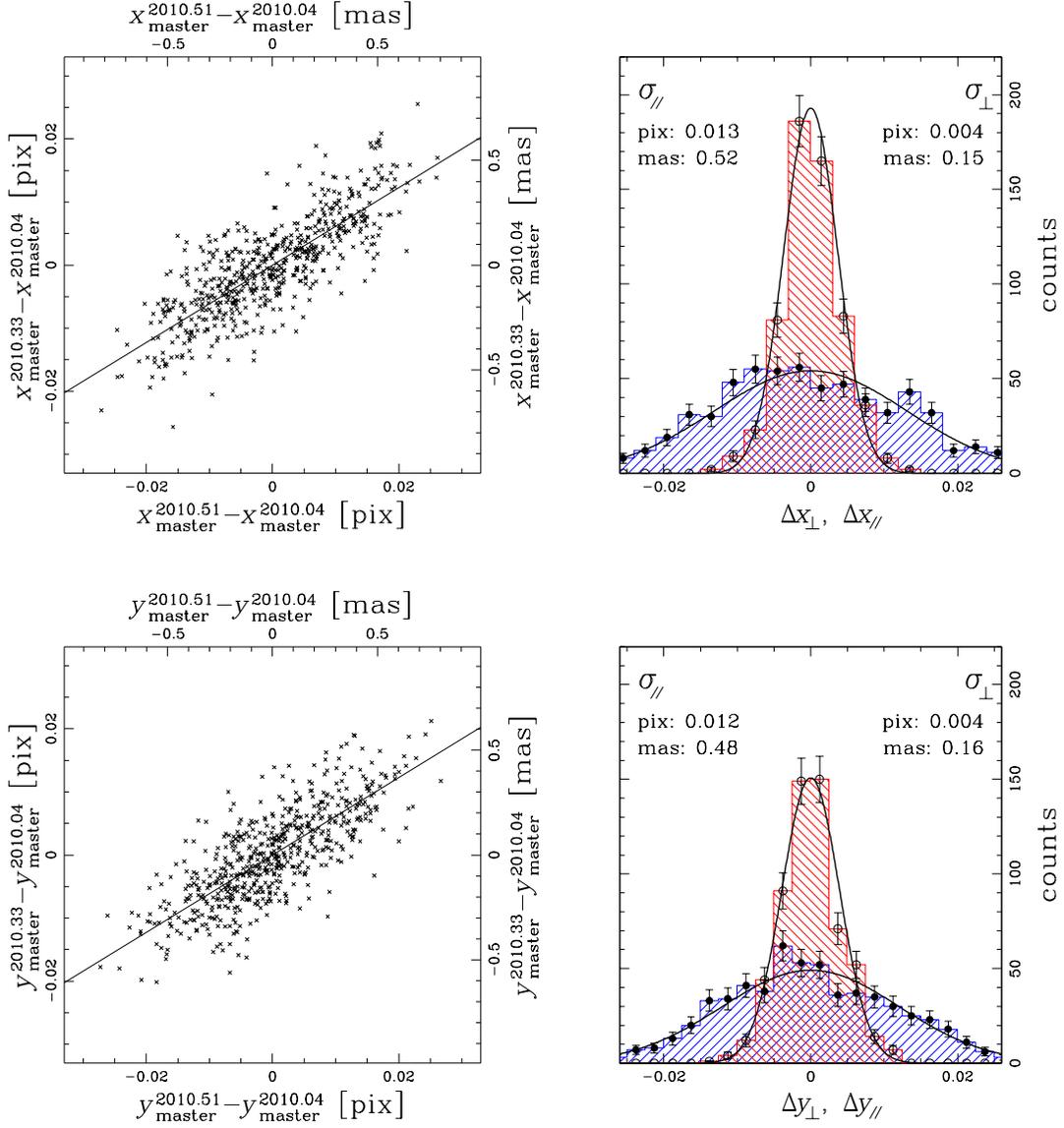}
\caption{\textit{Top-left:} correlation between $x$-positional
                 displacements as measured in three different 
                 epochs ($E1$, $t_1$=2010.04; $E2$, $t_2$=2010.33; 
                 $E3$, $t_3$=2010.51) for the F814W exposures. 
                 The slope of the straight line is computed as the the
                 ratio between the two time baselines of $E2-E1$ and
                 $E3-E1$. 
         \textit{Top-right:} histograms of the displacement 
                 distributions along (blue) and perpendicular 
                 to (red) the straight line. 
         \textit{Bottom panels:} the same for $y$-positional
                 displacements. See the text for details.}
\label{fig:ipm}
\end{figure*}

The globular cluster $\omega$~Cen has the largest
internal-proper-motion dispersion among Galactic globular clusters
despite being more than two times farther than the closest ones (4.7
kpc, van der Marel \& Anderson 2010, versus 2.2 kpc for M~4 and 2.3
kpc for NGC~6397, Harris 1996, Dec. 2010 revision).  In this section
we show that, by applying our GD solution, we are able to measure this
dispersion in just a few months.
 
To minimize chromatic effects, we used only F814W images and created
three distortion-free frames, one for each of the three available
epochs ($t_1$=2010.04, $t_2$=2010.33, $t_3$=2010.51, hereafter $E1$,
$E2$, $E3$ respectively).  We selected well-exposed, unsaturated stars
(instrumental magnitude $-13.5<{\rm F814W}<-12.5$) in common to all
the three epochs, and compared the positional displacements seen
between $(E3-E1)$ and $(E2-E1)$.

Detection of the intrinsic motion of the stars would reveal itself as
a correlation between the displacements $(E3-E1)$ and $(E2-E1)$,
proportional to the ratio of the respective time baselines
$(t_2-t_1)/(t_3-t_1)$.  On the contrary, had these displacements been
dominated by random errors, we would expect no correlation at all
between the two displacements, or at least a correlation with a
different slope.

This is shown on the left panels of Fig.~\ref{fig:ipm}, where we see a
clear correlation between the two coupled epochs in both the $x$- and
$y$-directions (top and bottom panels, respectively).  Note that the
slope of the straight line simply corresponds to the ratio between the
two time base-lines and it is not a fit to the data points.  If we
assume a Gaussian distribution for the observed displacements, the
dispersions along the line, and those perpendicular to it, can give us
an estimate of the intrinsic proper-motion and the errors' dispersion,
respectively.  In the following we will derive a crude estimate for
both.

We used the symbols $\sigma_{31}$ and $\sigma_{21}$ [in mas] to
indicate the standard deviation of the displacements $(E3-E1)$ and
$(E2-E1)$.  We then considered that the \textit{observed}
proper-motion dispersion, $\sigma_{\rm obs}$, is related to the
intrinsic proper-motion dispersion ($\sigma_{\rm intr}$) and the
measurement errors ($\sigma_{\rm err}$) by the following equation:
$$ 
     \sigma_{\rm obs} = \sqrt{\sigma_{\rm intr}^2 + \sigma_{\rm err}^2}
                           ~~ [\textrm{mas yr}^{-1}].
$$ 
Therefore $\sigma_{\rm intr}$ can be obtained as:
\begin{equation}
\label{eq1}
\sigma_{\rm intr} = \sqrt{\sigma_{\rm obs}^2 - \sigma_{\rm err}^2} ~~
      [\textrm{mas yr}^{-1}].
\end{equation}

The quantity $\sigma_{\rm obs}$ can also be computed as the ratio
between the displacements' dispersions and their time baseline:
      \footnote{This assumes the errors within epochs $E3$ and $E2$ 
                to be similar, a reasonable working hypothesis. } 
\begin{equation}
     \label{eq2}
     \sigma_{\rm obs} = \frac{\sigma_{31}}{t_3-t_1} =
                        \frac{\sigma_{21}}{t_2-t_1} 
                        ~~ [\textrm{mas yr}^{-1}],
\end{equation}
from which follows the relation:
$$ 
     \sigma_{21} = \frac{t_2-t_1}{t_3-t_1}\sigma_{31}~~[\textrm{mas}].
$$

We defined the dispersion of data points along the line in
Fig.\ref{fig:ipm} as $\sigma_{\parallel}$, which is an observable
quantity and related to the others by the equation
$$ \sigma_{\parallel} = \sqrt{\sigma_{31}^2+\sigma_{21}^2} =
\sqrt{\sigma_{31}^2+{\frac{(t_2-t_1)^2}{(t_3-t_1)^2}}\sigma_{31}^2} =
\sigma_{31} \sqrt{ 1+{\frac{(t_2-t_1)^2}{(t_3-t_1)^2}} }
~~[\textrm{mas}],
$$ which implies
$$ \sigma_{31} =
\frac{\sigma_{\parallel}}{\sqrt{1+{\frac{(t_2-t_1)^2}{(t_3-t_1)^2}}}}
~~[\textrm{mas}].
$$ The latter, substituted in Eq.~\ref{eq2}, gives
\begin{equation}
\label{eq3}
\sigma_{\rm obs} =
\frac{\sigma_{\parallel}}{(t_3-t_1)\sqrt{1+{\frac{(t_2-t_1)^2}{(t_3-t_1)^2}}}}
~ ~ {\rm [mas ~ yr^{-1}]},
\end{equation}
which relates the observed proper-motion dispersion to the observed
dispersion along the expected correlation.  Assuming any deviation
from the line to represents the total error of the data point, an
estimate of $\sigma_{\rm err}$ can be obtained by the dispersion of
data points perpendicularly to the line:
\begin{equation}
\label{eq4}
\sigma_{\rm err} =
\frac{\sigma_{\perp}}{(t_3-t_1)\sqrt{1+{\frac{(t_2-t_1)^2}{(t_3-t_1)^2}}}}
~ ~ {\rm [mas ~ yr^{-1}]},
\end{equation}

Top-right panel of Fig.~\ref{fig:ipm} shows the histograms of the
$x$-displacement dispersions along (filled circles) and perpendicular
(open circles) to the expected correlations.  Poisson error bars and
Gaussian best fits are also shown.  Standard deviations
$\sigma_{\parallel}$ and $\sigma_{\perp}$ are computed as the
68.27$^{\rm th}$ percentile of the distribution residuals around their
median value.  The bottom-right panel shows the same for $y$-axis
displacement dispersions.

\begin{figure*}[t!]
\centering
\includegraphics[width=14cm]{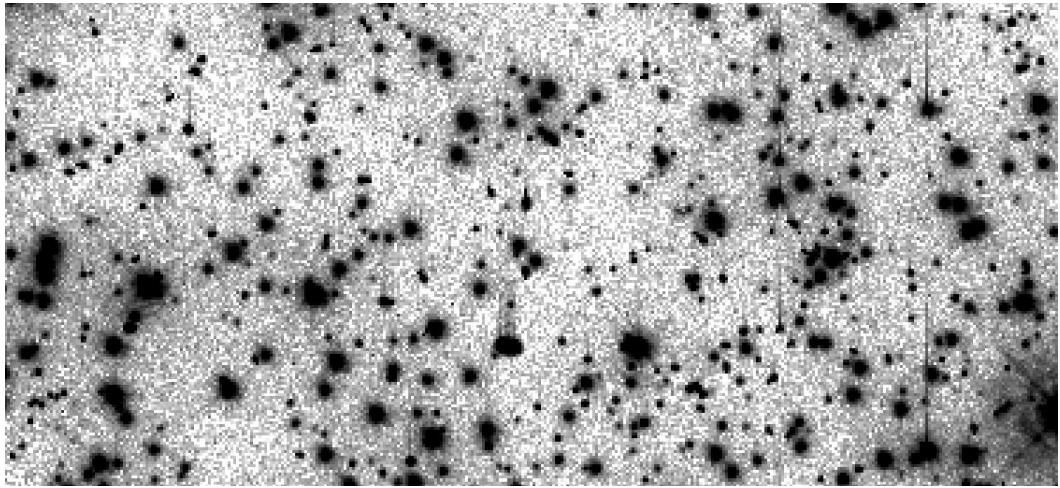}
\caption{A region of $350\times150$ pixels far from the readout
  register of a F275W exposure (900 s) obtained in July 2010
  (\texttt{ibc308yeq\_flt}), where charge-transfer efficiency problems
  are already visible.}
\label{fig:cti}
\end{figure*}

Taking the average of $\sigma_{\parallel}$ and $\sigma_{\perp}$ from
the displacements along $x$ and $y$, we had
$\sigma_{\parallel}\simeq0.50$ mas and $\sigma_{\perp}\simeq 0.15$
mas. By using the previous equations~\ref{eq1}, ~\ref{eq3} and
\ref{eq4}, we obtained an intrinsic dispersion of $\sigma_{\rm intr} =
0.86 \pm 0.27$ mas yr$^{-1}$, which is in remarkable agreement with
the value $\sigma_{\mathrm{1-D}} = 0.83\pm0.07$ mas yr$^{-1}$ measured
by Anderson \& van der Marel (2010) using a time baseline almost 9
times larger (4.07 years vs.\ 172 days!), and more images.  This result
is even more astonishing if we consider that we did not use local
transformations (see, e.g., Bedin et al.\ 2003; Anderson et al.\ 2006,
Anderson \& van der Marel 2010) which would further reduce systematic
residuals in the GD solution, as well as any correction for breathing
(which can introduce small low-order terms) or charge-transfer
inefficiency (which is already plaguing this new camera; see
Figure~\ref{fig:cti}).

As a final external check on the achieved accuracy, we can assume the
uncertainties on $E1$ to cancel out and $E2$ and $E3$ to equally
contribute to $\sigma_\perp$
($=\sqrt{\sigma_{E2}^2+\sigma_{E3}^2}=\sqrt{2}\times\sigma_{E}$).
Having $\sim$9 exposures per epoch, we can infer a 1-D positional
accuracy of 0.008 pixels, $\sim$0.3 mas
($\sqrt{9-1}\times\sigma_\perp/\sqrt{2}$), which is consistent with
the value reported in Fig.~\ref{fig:res}.  It should also be noted
that, at this level of accuracy, there is a considerable interplay
between the derived PSF models and the GD solution, which might play
some role on the achievable astrometric precision.

\section{Putting the chip-based solution into a global framework.}
\label{sec:global} 

On account of breathing and other phenomena, the distortion solution
is not perfectly stable over time.  The impact of breathing is not 
well enough understood to predict its impact on the distortion, so
the best we can hope to achieve is an average distortion solution 
and a sense of how stable the solution is about this average.  It 
is typically the low-order terms that are the most time-variable, so
it made sense above to construct the best single-chip-based solution
first and only later to consider the less accurate larger-scale terms
that relate the two chips to a common frame.  We do that here with the 
understanding, that for the highest-precision differential astrometry, 
it is always best to perform the measurements as locally as possible, 
provided that there are an adequate number of reference stars.  Some 
projects, however have few reference objects and require knowledge 
of the distortion over the full extent of the field of view (FOV).

We solved for these global terms in several stages.  First, we
put the two chips into a meta coordinate frame.  Next, we solved 
for the linear skew terms present in the combined system.  Then 
we solved for the overall scaling, so that all filters would 
have the same scaling, in terms of pixels per arcsecond.  Finally, 
we solved for the positional offset between the filters.

\subsection{The Meta Coordinate Frame}
\label{ssec:global} 

We based our meta-coordinate frame on the distortion-corrected
frame of the bottom chip (chip[1]).  The bottom chip is already 
in this system, so we simply needed to find the transformation that
put the top chip (chip[2]) into this same system.  Since there 
is a gap of about 35 pixels between the frames, the two chips 
of a single exposure naturally have no sources in common.  This 
means that we will have to use an intermediate set of observations 
to accomplish the mapping.

We took all the F606W observations and identified all pairs of
observations that had either an offset of more than 500 pixels or an
orientation difference of more than 10 degrees.  (If the pointings are
too similar then the overlap will not be sufficient.)  Since we wanted
as many different pairs of exposures as possible, we also incorporated
images from PID-12094 (PI-Kozhurina-Platais), which provides
additional roll angles.  There were a total of 32 variously
overlapping F606W exposures, so in general we could work with 992
overlapping pairs.

For each qualifying pair, we found which chip(s) in the first 
exposure had significant overlap with both chips in the second 
exposure.  We first corrected all positions for distortion using 
the solution found above.  We next used the positions of common
stars to solve for the 4-parameter conformal linear transformation 
from the top chip of the second exposure into the overlapping chip
of the first exposure.  We then solved for the analogous
transformation from the overlapping chip to the bottom chip of 
the second exposure.  By combining these two linear transformations,
we were able to bootstrap positions in the top chip to positions 
in the bottom chip.  

Using all pairs of exposures with good intermediate-chip overlap, 
we found a transformation of the form:
\begin{eqnarray}
\begin{array}{l}
xc_1 = xc_2 + \Delta x + A \phi_x + B \phi_y \\  
yc_1 = yc_2 + \Delta y + C \phi_x + D \phi_y,
\end{array}
\end{eqnarray}

where $\phi_x = (xc_2-2048)/4096$ and $\phi_y = (yc_2-1024)/2048$
to relate the distortion-corrected coordinates in the top chip 
$(xc_2,yc_2)$ into the system of the bottom frame.  This was found
for each exposure where both of its chips overlapped significantly
with a chip from a different exposure.

We then found average values for the six parameters.  This gave
us a rough meta-frame system.  Since the single-chip overlaps 
can be limited, we then iterated this solution, this time using
both chips in the comparison exposure (by means of the meta solution) 
to examine any residuals in the positioning of the top chip 
relative to the bottom.  We converged upon:
\begin{eqnarray}
\begin{array}{l}
xc_1 = xc_2 -    2.020 - 12.035 \phi_x -  2.367 \phi_y  \\
yc_1 = yc_2 + 2073.410 +  2.363 \phi_x - 12.031 \phi_y, 
\end{array}
\end{eqnarray}
for the F606W data set.  This means that the central pixel of 
the top chip (2048,1024), where $\phi_x$ and $\phi_y$ are zero, 
is located at coordinate $(2048 - 2.02, 1024 + 2073.41)$ = 
$(2045.98, 3097.41)$ in the distortion-corrected frame of the 
bottom chip.  The near-conformal nature of this transformation
($A \simeq D$ and $B \simeq -C$) demonstrates that the skew
was accurately measured in the original solution.

Now that we had a master frame for the F606W exposures, we used 
the F606W data set as comparison images to solve for the same parameters
for the other filters.  We would expect them to be similar, but 
since the solution for each filter was found independent of the 
other filters, there could be small scale or orientation changes 
that would impact the chip[1] frame, and hence the location of
chip[2] in chip[1] coordinates.  Table~\ref{tab3} provides these 
inter-chip transformation parameters for all filters.

\begin{table*}[th!]
\caption{Coefficients used to map the coordinate system of 
         chip[2] into the frame of chip[1] for each filter.  }
\label{tab3}
\centering
\begin{tabular}{c|rrrrrr|r|rr}
\hline
\hline
\multicolumn{1}{c|}{FILT} & 
\multicolumn{1}{c}{$\Delta x$} & 
\multicolumn{1}{c}{$\Delta y$} & 
\multicolumn{1}{c}{$A$} & 
\multicolumn{1}{c}{$B$} & 
\multicolumn{1}{c}{$C$} & 
\multicolumn{1}{c}{$D$} &
\multicolumn{1}{|c}{SCALE} &
\multicolumn{1}{|c}{DX} &
\multicolumn{1}{c}{DY} \\
\hline
F225W  & $-$2.024 & 2073.362 & $-$12.152 &  $-$2.355 &   2.306 & $-$12.146 &  1.000110 &  0.200 &  0.405 \\
F275W  & $-$2.001 & 2073.354 & $-$12.123 &  $-$2.318 &   2.310 & $-$12.116 &  0.999997 &  0.265 &  0.120 \\
F336W  & $-$2.023 & 2073.383 & $-$12.068 &  $-$2.384 &   2.367 & $-$12.068 &  0.999851 &  0.440 & $-$0.495 \\
F390W  & $-$2.001 & 2073.411 & $-$12.098 &  $-$2.391 &   2.397 & $-$12.110 &  0.999894 &  0.280 & $-$0.455 \\
F438W  & $-$2.003 & 2073.402 & $-$12.097 &  $-$2.368 &   2.367 & $-$12.096 &  1.000007 & $-$0.075 & $-$0.510 \\
F555W  & $-$1.996 & 2073.434 & $-$12.063 &  $-$2.356 &   2.360 & $-$12.046 &  1.000121 & $-$0.085 & $-$0.535 \\
F606W  & $-$2.020 & 2073.410 & $-$12.035 &  $-$2.367 &   2.363 & $-$12.031 &  0.999997 &  0.000 &  0.000 \\
F775W  & $-$1.975 & 2073.409 & $-$11.995 &  $-$2.346 &   2.339 & $-$11.987 &  1.000004 &  0.260 & $-$1.470 \\
F814W  & $-$2.048 & 2073.463 & $-$12.038 &  $-$2.384 &   2.387 & $-$12.033 &  1.000048 & $-$0.070 & $-$0.390 \\
F850LP & $-$1.969 & 2073.389 & $-$12.027 &  $-$2.299 &   2.300 & $-$12.017 &  1.000158 &  0.155 & $-$0.725 \\ 
\hline
\hline
\end{tabular}
\end{table*}

\subsection{Solving for the relative scale for each filter.}
\label{ssec:scale_by_filt} 

The solution for each chip/filter combination described above was 
performed without reference to other chips and filters.  The goal
of this section is to tie everything together into a common system.
In the coordinate transformations above, we always solved for a 
scale before we examined residuals.  The reason for this is that 
the scale of the telescope is always changing, partly due to breathing 
and partly due to velocity aberration.

Now that the offset and rotation for each chip/filter combination
had been determined to place each of the chips into a common master 
frame, the overall scale was the last of the linear parameters
to be solved for.  To do this, we constructed a master frame based
on only the F606W exposures using only transformations allowed for
offset and rotation, but no scale changes.  This way, the frame 
would represent the average scale of the F606W exposures.

\begin{figure*}[t!]
\centering
\includegraphics[width=14cm]{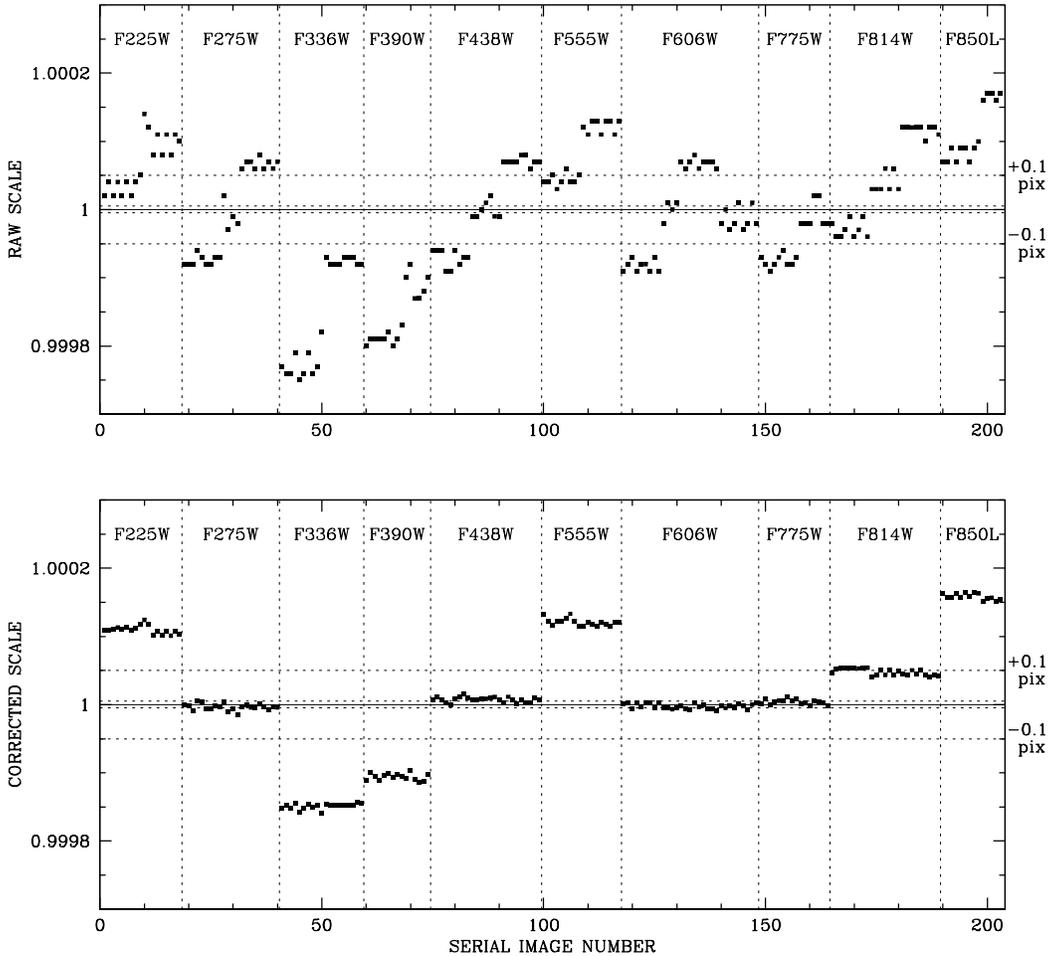}
\caption{\textit{Top:}    The solved-for scale difference between 
                  the F606W-based master frame and each individual
                  exposure.  The 203 exposures are ordered by filter.
                  The dotted lines correspond to scale changes that
                  would expand the field of view by $\pm$0.01 and
                  $\pm$0.10 pixel.
         \textit{Bottom:} The same points, but after velocity
                  aberration has been accounted for.}
\label{fig:scale_by_filt}
\end{figure*}

We then transformed the positions of the stars in each of the
exposures for each filter (203 in total, including the PID-12094 data)
into this reference frame and took note of the scale factor of the
linear transformation.  We plot this scale factor for each exposure on
the top panels of Figure~\ref{fig:scale_by_filt}.  The images are
ordered by filter and the filters are separated by a vertical dotted
line.  It is clear that there is some trend with filter, but there is
considerable intra-filter scatter as well.  This scatter could be due
to velocity aberration or breathing.

We divided each of the above scale measurements by the
\texttt{VAFACTOR} keyword, taken from the image header.  It reports
the expected special-relativistic variation in the plate scale, which
is related to the dot-product between the average telescope velocity
vector during the exposure and the direction to the target (see Cox \&
Gilliland 2002).  After making this deterministic correction, we found
that the global exposures for each filter were now consistent to
within 0.01 pixel (as shown on the bottom panels of
Fig.~\ref{fig:scale_by_filt}).  We solved for the average scaling for
each filter and included this correction in the meta distortion
solution.  The average scaling we found for each filter is given in
the eighth column of Table~\ref{tab3}.  They are in good qualitative
agreement with the findings in Figure 2 of Kozhurina-Platais et
al.\ (2010).  We note that the linear terms of UVIS appear to be
considerably more stable than those of ACS/WFC.  Figure 6 of Anderson
(2006) shows that the ACS FOV changes in radius by $\pm$0.03 pixel due
to breathing, while the residuals here are about 0.01 pixel.

We want emphasize that breathing can introduce errors larger than
0.008 pixel into the distortion solution. Our single-chip-based  GD
correction was generated comparing largely half-chip to half-chip
positions (due to the dithering scheme of the exposures). Therefore,
the global effect due to breathing is reduced to about a quarter with
respect to what it would be by comparing whole chips to whole chips
(since errors tend to be quadratic).

\subsection{Solving for the offset for each filter.}
\label{ssec:offset_by_filt} 

Wedge effects can cause individual filters to induce small shifts 
in the positions of stars on the chip.  Since the zeropoint of our 
distortion solution for each filter is entirely arbitrary, we have 
the flexibility of choosing it in a convenient way.  We would like 
the distortion solution to have the property that, if a star is observed 
through two different filters with no dither, the two filters will 
report the same distortion-corrected position for the star, even if 
the wedge effect may cause the apparent positions on the chip to differ 
by more than a pixel.

We identified two sets of consecutive exposures of the $\omega$~Cen
central field that had no commanded \texttt{POS-TARGs} between them
and found the offset for each filter that best registered its
distortion-corrected positions with those of the F606W images.  These
offsets are given in the last two columns of Table~\ref{tab3}.  Based
on the agreement between the two comparisons we have for each filter,
these offsets should be accurate to about 0.05 pixel.  Note that the
offset for F775W is quite large (1.47 pixel); the others are typically
0.5 pixel.

\subsection{The absolute scale.}
\label{ssec:absolute_scale}

The distortion solution presented here was constructed to match
the pixel scale and orientation of the center of chip[1].  To determine
the absolute scale of this frame, we compared the commanded offsets
in arcseconds (from the commanded {\tt POS\_TARGs}) with the achieved 
offsets (in pixels).  

PID-11911 contained visits in which 9 exposures in F606W were taken in
a 3$\times$3 grid with offsets of 40\arcsec\ between the exposures.
We measured the achieved offsets from the central exposure in
distortion-corrected pixels and found that the grid spacing in pixels
corresponded to 1005.67 pixels in one visit and 1005.79 in the second
visit.  This implies a plate scale for our frame of $25.143 \pm 0.002$
pixels per arcsecond, which corresponds to $39.773\pm0.003$ mas
pixel$^{-1}$. This value is consistent with the independent estimate
of the scale given in Paper~I ($39.770\pm0.01$ mas, internal errors
only), which is based on the best knowledge of the ACS/WFC absolute
scale.

\subsection{The final meta frame}
\label{ssec:final_meta}

After establishing all of these global parameters as described, we noticed 
that the frame we had adopted, which was centered on the central pixel 
of chip[1] (the bottom chip), extended into negative coordinates in the 
lower left of the frame.  This is due to our choice of center and the 
nature of the intrinsic skew present in the detector.  Since it is often 
convenient to work with positive coordinates, we decided to add 200 
to the meta coordinate system in $y$, so that positions within the
detector would always have positive coordinate values.

\section{Conclusions}
\label{sec:conc} 

By using a large number of well dithered data-sets taken with
different roll angles, we have modeled the GD of WFC3/UVIS by means of
a self-calibration.  The solution is an improvement with respect to
Paper~I and consists of a set of third-order-correction coefficients
plus a finely-spaced look-up table of residuals for each chip in 10
filters, namely:\ F225W, F275W, F336W, F390W, F438W, F555W, F606W,
F775W, F814W and F850LP.  The hybrid solution has been shown to
correct the manufacturing defects and the high-spatial frequency
residuals seen in Paper~I.

The use of these corrections removes the distortion over the entire
area of each chip to an accuracy significantly better than $\sim$0.01
pixel (i.e. better than $\sim$0.4 mas).  As a demonstrative test we
applied our solution to F814W exposures collected at different epochs
and were able to measure the internal motion of $\omega$~Cen in just
few months, finding values consistent with the most recent
determinations.

The initial solution was constructed for each chip for each filter
independently.  This is because we did not want any possible motion
between the chips or large-scale breathing non-linearities to impact
our solution.  Once we had the solution for each filter and chip, we
unified them into a single common coordinate system.  We found the
linear transformation that took the top chip into the frame of the
bottom chip for each filter, we found the relative scalings between
the different filters and finally we found the offsets caused by the
wedge-effect of each filter.

We make our \texttt{FORTRAN} routine \texttt{WFC3\_UVIS\_gc.f}
publicly available (at the url
\texttt{www\-.stsci\-.edu\-/$\sim$jayander/WFC3UV\_GC/}) to the
astronomical community.  It requires 4 quantities in input: the raw
positions $x_{\rm raw}$ and $y_{\rm raw}$, the chip number and the
filter. In output it produces corrected positions $x_{\rm corr}$ and
$y_{\rm corr}$ in the meta-coordinate frame.

In addition to the \texttt{FORTRAN} code, we will also provide the
solution in the form of simple fits images.  The ``image\_format''
directory of the website above contains forty images, one for each
coordinate/chip/filter combination.  As an example, the image {\tt
  gc\_wfc3uvis\_f606w\_chip1x.fits} is a 4096$\times$2051 real$\ast$4
image with each pixel in the image giving the $x$ coordinate of the
distortion-corrected position of that pixel.  To distortion-correct
decimal pixel locations, the image can simply be interpolated
bi-linearly.  These images should make our solution accessible for
those who use languages other than \texttt{FORTRAN}.

Finally, we want to state clearly to the reader that this is not the
official GD calibration.  The IDCTAB file, which contains independent
polynomial calibrations for the 10 UVIS filters, is installed in
STScI's Calibration Database System and is used for the OPUS pipeline
processing of rectified \texttt{DRZ} images (Kozhurina-Platais et
al.\ 2009, 2010).  The two solutions were designed for different
goals.  Ours was constructed with high-precision differential
astrometry in mind, while the official solution was more focused on
the absolute transformation onto the focal plane.

\acknowledgements We thank Alceste Z.\ Bonanos for polishing the
manuscript.  A.B.\ acknowledges the support by the STScI under the
2008 graduate research assistantship program and by MIUR under program
PRIN2007 (prot.\ 20075TP5K9).

\newpage
\clearpage
\begin{center}
\textbf{ A. Appendix}
\end{center}
\appendix
\label{appA}

Table~A lists the polynomial coefficients for all the 10
filters (F225W, F275W and F336W coefficients are from Bellini \& Bedin
2009, reported here for completeness).
\label{tabA}
\footnotesize{
\begin{center}
\begin{longtable}{rcrrrrrcrrrr}
\multicolumn{12}{l}{Table A: Coefficients of the 3$^{\rm rd}$-order polynomial solution
  for chip[1] and chip[2] for all the 10 filters.}\\
\hline
\hline
\multicolumn{1}{c}{Term $\!(w)\!\!\!\!\!$}&Polyn.$\!\!\!\!\!\!\!\!\!$&$a_{w,[1]}$&$b_{w,[1]}$&$a_{w,[2]}$&$b_{w,[2]}$&
\multicolumn{1}{c}{$\phantom{sp}$Term $\!(w)\!\!\!\!\!$}&Polyn.$\!\!\!\!\!\!\!\!\!$&$a_{w,[1]}$&$b_{w,[1]}$&$a_{w,[2]}$&$b_{w,[2]}$\\
\hline
\endfirsthead
\multicolumn{12}{r}{{\footnotesize{Continues from previous page}}}\\
\hline
\hline
\multicolumn{1}{c}{Term $\!(w)\!\!\!\!\!$}&Polyn.$\!\!\!\!\!\!\!\!\!$&$a_{w,[1]}$&$b_{w,[1]}$&$a_{w,[2]}$&$b_{w,[2]}$&
\multicolumn{1}{c}{$\phantom{sp}$Term $\!(w)\!\!\!\!\!$}&Polyn.$\!\!\!\!\!\!\!\!\!$&$a_{w,[1]}$&$b_{w,[1]}$&$a_{w,[2]}$&$b_{w,[2]}$\\
\hline
\endhead
\hline
\multicolumn{12}{r}{{\footnotesize{Continues on next page}}}\\
\endfoot
\hline
\hline
\endlastfoot
&&\multicolumn{4}{c}{\textbf{F225W}}&&&\multicolumn{4}{c}{\textbf{F275W}}\\
1&$\tilde{x}$&            $  0.000\!$&$\!129.230\!$&$  0.000\!$&$\!140.270\!$&1&$\tilde{x}$&            $  0.000\!$&$\!129.270\!$&$  0.000\!$&$\!140.285\!$\\
2&$\tilde{y}$&            $  0.000\!$&$\!  1.935\!$&$  0.000\!$&$\! -4.215\!$&2&$\tilde{y}$&            $  0.000\!$&$\!  1.925\!$&$  0.000\!$&$\! -4.221\!$\\
3&$\tilde{x}^2$&          $ 12.120\!$&$\!  0.591\!$&$ 12.021\!$&$\!  0.773\!$&3&$\tilde{x}^2$&          $ 12.102\!$&$\!  0.581\!$&$ 12.016\!$&$\!  0.781\!$\\
4&$\tilde{x}\tilde{y}$&   $ -6.279\!$&$\!  5.553\!$&$ -6.057\!$&$\!  5.496\!$&4&$\tilde{x}\tilde{y}$&   $ -6.284\!$&$\!  5.547\!$&$ -6.040\!$&$\!  5.493\!$\\
5&$\tilde{y}^2$&          $  0.064\!$&$\! -3.227\!$&$  0.001\!$&$\! -3.058\!$&5&$\tilde{y}^2$&          $  0.061\!$&$\! -3.241\!$&$  0.001\!$&$\! -3.048\!$\\
6&$\tilde{x}^3$&          $  0.176\!$&$\!  0.029\!$&$  0.149\!$&$\!  0.156\!$&6&$\tilde{x}^3$&          $  0.178\!$&$\!  0.033\!$&$  0.144\!$&$\!  0.163\!$\\
7&$\tilde{x}^2\tilde{y}$& $ -0.057\!$&$\!  0.033\!$&$  0.022\!$&$\! -0.009\!$&7&$\tilde{x}^2\tilde{y}$& $ -0.056\!$&$\!  0.054\!$&$  0.026\!$&$\!  0.007\!$\\
8&$\tilde{x}\tilde{y}^2$& $  0.004\!$&$\! -0.041\!$&$  0.061\!$&$\! -0.026\!$&8&$\tilde{x}\tilde{y}^2$& $  0.005\!$&$\! -0.041\!$&$  0.051\!$&$\! -0.025\!$\\
9&$\tilde{y}^3$&          $  0.035\!$&$\! -0.023\!$&$  0.032\!$&$\!  0.028\!$&9&$\tilde{y}^3$&          $  0.033\!$&$\! -0.012\!$&$  0.032\!$&$\!  0.020\!$\\
&&\multicolumn{4}{c}{\textbf{F336W}}&&&\multicolumn{4}{c}{\textbf{F390W}}\\
1&$\tilde{x}$&           $  0.000\!$&$\!129.438\!$&$  0.000\!$&$\!140.315\!$&1&$\tilde{x}$&           $  0.000\!$&$\!129.364\!$&$  0.000\!$&$\!140.216\!$\\
2&$\tilde{y}$&           $  0.000\!$&$\!  1.786\!$&$  0.000\!$&$\! -4.322\!$&2&$\tilde{y}$&           $  0.000\!$&$\!  1.866\!$&$  0.000\!$&$\! -4.258\!$\\
3&$\tilde{x}^2$&         $ 12.091\!$&$\!  0.676\!$&$ 11.994\!$&$\!  0.672\!$&3&$\tilde{x}^2$&         $ 12.058\!$&$\!  0.664\!$&$ 11.959\!$&$\!  0.720\!$\\
4&$\tilde{x}\tilde{y}$&  $ -6.188\!$&$\!  5.565\!$&$ -6.135\!$&$\!  5.476\!$&4&$\tilde{x}\tilde{y}$&  $ -6.199\!$&$\!  5.551\!$&$ -6.114\!$&$\!  5.472\!$\\
5&$\tilde{y}^2$&         $  0.065\!$&$\! -3.155\!$&$  0.004\!$&$\! -3.152\!$&5&$\tilde{y}^2$&         $  0.067\!$&$\! -3.161\!$&$ -0.007\!$&$\! -3.116\!$\\
6&$\tilde{x}^3$&         $ -0.062\!$&$\!  0.004\!$&$ -0.151\!$&$\!  0.189\!$&6&$\tilde{x}^3$&         $  0.042\!$&$\!  0.014\!$&$  0.002\!$&$\!  0.177\!$\\
7&$\tilde{x}^2\tilde{y}$&$ -0.097\!$&$\!  0.034\!$&$  0.074\!$&$\! -0.027\!$&7&$\tilde{x}^2\tilde{y}$&$ -0.083\!$&$\!  0.026\!$&$  0.070\!$&$\! -0.017\!$\\
8&$\tilde{x}\tilde{y}^2$&$  0.016\!$&$\! -0.061\!$&$  0.040\!$&$\!  0.005\!$&8&$\tilde{x}\tilde{y}^2$&$  0.016\!$&$\! -0.051\!$&$  0.074\!$&$\!  0.001\!$\\
9&$\tilde{y}^3$&         $  0.033\!$&$\!  0.016\!$&$  0.033\!$&$\!  0.014\!$&9&$\tilde{y}^3$&         $  0.050\!$&$\! -0.014\!$&$  0.055\!$&$\!  0.022\!$\\
&&\multicolumn{4}{c}{\textbf{F438W}}&&&\multicolumn{4}{c}{\textbf{F555W}}\\
1&$\tilde{x}$&           $  0.000\!$&$\!129.378\!$&$  0.000\!$&$\!140.276\!$&1&$\tilde{x}$&           $  0.000\!$&$\!129.357\!$&$  0.000\!$&$\!140.292\!$\\
2&$\tilde{y}$&           $  0.000\!$&$\!  1.897\!$&$  0.000\!$&$\! -4.258\!$&2&$\tilde{y}$&           $  0.000\!$&$\!  1.931\!$&$  0.000\!$&$\! -4.185\!$\\
3&$\tilde{x}^2$&         $ 12.035\!$&$\!  0.611\!$&$ 11.974\!$&$\!  0.771\!$&3&$\tilde{x}^2$&         $ 12.146\!$&$\!  0.579\!$&$ 12.038\!$&$\!  0.770\!$\\
4&$\tilde{x}\tilde{y}$&  $ -6.234\!$&$\!  5.539\!$&$ -6.069\!$&$\!  5.492\!$&4&$\tilde{x}\tilde{y}$&  $ -6.226\!$&$\!  5.522\!$&$ -6.037\!$&$\!  5.464\!$\\
5&$\tilde{y}^2$&         $  0.063\!$&$\! -3.161\!$&$  0.002\!$&$\! -3.079\!$&5&$\tilde{y}^2$&         $  0.062\!$&$\! -3.236\!$&$ -0.005\!$&$\! -3.034\!$\\
6&$\tilde{x}^3$&         $  0.116\!$&$\!  0.030\!$&$  0.061\!$&$\!  0.167\!$&6&$\tilde{x}^3$&         $  0.304\!$&$\!  0.032\!$&$  0.229\!$&$\!  0.119\!$\\
7&$\tilde{x}^2\tilde{y}$&$ -0.051\!$&$\!  0.054\!$&$  0.043\!$&$\!  0.027\!$&7&$\tilde{x}^2\tilde{y}$&$ -0.043\!$&$\!  0.047\!$&$ -0.014\!$&$\!  0.015\!$\\
8&$\tilde{x}\tilde{y}^2$&$  0.031\!$&$\! -0.033\!$&$  0.073\!$&$\! -0.022\!$&8&$\tilde{x}\tilde{y}^2$&$  0.023\!$&$\! -0.010\!$&$  0.054\!$&$\! -0.019\!$\\
9&$\tilde{y}^3$&         $  0.041\!$&$\! -0.022\!$&$  0.056\!$&$\!  0.023\!$&9&$\tilde{y}^3$&         $  0.034\!$&$\!  0.005\!$&$  0.054\!$&$\!  0.018\!$\\
&&\multicolumn{4}{c}{\textbf{F606W}}&&&\multicolumn{4}{c}{\textbf{F775W}}\\
1&$\tilde{x}$&           $  0.000\!$&$\!129.384\!$&$  0.000\!$&$\!140.280\!$&1&$\tilde{x}$&           $  0.000\!$&$\!129.322\!$&$  0.000\!$&$\!140.167\!$\\
2&$\tilde{y}$&           $  0.000\!$&$\!  1.868\!$&$  0.000\!$&$\! -4.251\!$&2&$\tilde{y}$&           $  0.000\!$&$\!  1.946\!$&$  0.000\!$&$\! -4.211\!$\\
3&$\tilde{x}^2$&         $ 12.044\!$&$\!  0.622\!$&$ 11.959\!$&$\!  0.762\!$&3&$\tilde{x}^2$&         $ 12.044\!$&$\!  0.593\!$&$ 12.022\!$&$\!  0.780\!$\\
4&$\tilde{x}\tilde{y}$&  $ -6.222\!$&$\!  5.526\!$&$ -6.069\!$&$\!  5.462\!$&4&$\tilde{x}\tilde{y}$&  $ -6.204\!$&$\!  5.538\!$&$ -6.040\!$&$\!  5.530\!$\\
5&$\tilde{y}^2$&         $  0.065\!$&$\! -3.211\!$&$  0.001\!$&$\! -3.063\!$&5&$\tilde{y}^2$&         $  0.062\!$&$\! -3.171\!$&$  0.005\!$&$\! -2.986\!$\\
6&$\tilde{x}^3$&         $  0.183\!$&$\!  0.026\!$&$  0.096\!$&$\!  0.156\!$&6&$\tilde{x}^3$&         $  0.273\!$&$\!  0.002\!$&$  0.139\!$&$\!  0.228\!$\\
7&$\tilde{x}^2\tilde{y}$&$ -0.054\!$&$\!  0.080\!$&$  0.037\!$&$\!  0.009\!$&7&$\tilde{x}^2\tilde{y}$&$ -0.052\!$&$\!  0.042\!$&$  0.123\!$&$\!  0.002\!$\\
8&$\tilde{x}\tilde{y}^2$&$  0.033\!$&$\! -0.034\!$&$  0.076\!$&$\! -0.020\!$&8&$\tilde{x}\tilde{y}^2$&$  0.059\!$&$\!  0.006\!$&$  0.020\!$&$\!  0.037\!$\\
9&$\tilde{y}^3$&         $  0.023\!$&$\!  0.009\!$&$  0.036\!$&$\!  0.049\!$&9&$\tilde{y}^3$&         $  0.075\!$&$\! -0.001\!$&$  0.032\!$&$\! -0.010\!$\\
&&\multicolumn{4}{c}{\textbf{F814W}}&&&\multicolumn{4}{c}{\textbf{F850L}}\\
1&$\tilde{x}$&           $  0.000\!$&$\!129.365\!$&$  0.000\!$&$\!140.168\!$&1&$\tilde{x}$&           $  0.000\!$&$\!129.318\!$&$  0.000\!$&$\!140.299\!$\\
2&$\tilde{y}$&           $  0.000\!$&$\!  1.897\!$&$  0.000\!$&$\! -4.275\!$&2&$\tilde{y}$&           $  0.000\!$&$\!  1.940\!$&$  0.000\!$&$\! -4.208\!$\\
3&$\tilde{x}^2$&         $ 12.025\!$&$\!  0.618\!$&$ 12.058\!$&$\!  0.765\!$&3&$\tilde{x}^2$&         $ 12.051\!$&$\!  0.547\!$&$ 11.981\!$&$\!  0.818\!$\\
4&$\tilde{x}\tilde{y}$&  $ -6.219\!$&$\!  5.516\!$&$ -6.084\!$&$\!  5.508\!$&4&$\tilde{x}\tilde{y}$&  $ -6.236\!$&$\!  5.522\!$&$ -5.985\!$&$\!  5.510\!$\\
5&$\tilde{y}^2$&         $  0.065\!$&$\! -3.190\!$&$  0.003\!$&$\! -3.089\!$&5&$\tilde{y}^2$&         $  0.058\!$&$\! -3.243\!$&$ -0.002\!$&$\! -2.992\!$\\
6&$\tilde{x}^3$&         $  0.192\!$&$\!  0.037\!$&$  0.064\!$&$\!  0.201\!$&6&$\tilde{x}^3$&         $  0.292\!$&$\!  0.024\!$&$  0.194\!$&$\!  0.150\!$\\
7&$\tilde{x}^2\tilde{y}$&$ -0.054\!$&$\!  0.072\!$&$  0.091\!$&$\!  0.022\!$&7&$\tilde{x}^2\tilde{y}$&$ -0.046\!$&$\!  0.065\!$&$  0.024\!$&$\!  0.027\!$\\
8&$\tilde{x}\tilde{y}^2$&$  0.047\!$&$\! -0.047\!$&$  0.073\!$&$\!  0.042\!$&8&$\tilde{x}\tilde{y}^2$&$  0.042\!$&$\! -0.050\!$&$  0.066\!$&$\! -0.030\!$\\
9&$\tilde{y}^3$&         $  0.037\!$&$\! -0.004\!$&$  0.027\!$&$\!  0.053\!$&9&$\tilde{y}^3$&         $  0.058\!$&$\!  0.001\!$&$  0.039\!$&$\!  0.040\!$\\
\end{longtable}
\end{center}}


\clearpage



\begin{thebibliography}{}

\bibitem[(1)]{1} Anderson, J., \& King, I.~R.\ 1999, \pasp, 111, 1095

\bibitem[(2)]{2} Anderson, J., \& King, I.~R.\ 2000, \pasp, 112,
  1360

\bibitem[(3)]{3} Anderson, J.\ 2002, The 2002 {\it HST} Calibration
  Workshop: Hubble after the Installation of the ACS and the NICMOS
  Cooling System, 13

\bibitem[(4)]{4} Anderson, J., \& King, I.~R.\ 2003, \pasp, 115, 113

\bibitem[(5)]{5} Anderson, J.\ 2006, The 2005 {\it HST} Calibration
  Workshop: Hubble After the Transition to Two-Gyro Mode, 11

\bibitem[(6)]{6} Anderson, J., \& King, I.~R.\ 2004, Instrument
  Science Report ACS 2004-15, 51 pages, 1

\bibitem[(7)]{7} Anderson, J., \& King, I.~R.\ 2006, Instrument
  Science Report ACS 2006-01, 34 pages, 1, AK06

\bibitem[(8)]{8} Anderson, J., Bedin, L.~R., Piotto, G., Yadav, R.~S.,
  \& Bellini, A.\ 2006, \aap, 454, 1029

\bibitem[(9)]{9} Anderson, J.\ 2007, Instrument Science Report ACS
  2007-08, 12 pages, 8

\bibitem[(10)]{10} Anderson, J., \& van der Marel, R.~P.\ 2010, \apj,
  710, 1032

\bibitem[(11)]{11} Bellini, A., \& Bedin, L.~R. 2009, \pasp, 121,
  1419, Paper~I

\bibitem[(12)]{12} Bellini, A., \& Bedin, L.~R. 2010, \aap, 517, A34

\bibitem[(13)]{13} Cox, C., \& Gilliland, R.~L.\ 2002, The 2002
  \textit{HST} Calibration Workshop : Hubble after the Installation of
  the ACS and the NICMOS Cooling System, 58

\bibitem[(14)]{14} Kozhurina et al.\ 2009, Instrument Science Report
  WFC3 2009-033, 22 pages, 33

\bibitem[(15)]{15} Kozhurina-Platais, V., Cox, C, Petro, L., Dulude,
  M., and Mack, J.  \textit{"Multi-Wavelength Geometric Distortion
    Solution for WFC3/UVIS and IR"} in Proceedings of the 2010 HST
  Calibration Workshop, Eds. S. Deustua and C. Oliveira (Baltimore:
  STScI), in press (2010).

\bibitem[(16)]{16} Harris, W.\ E.\ 1996, \aj, 112, 1487

\bibitem[(17)]{17} van der Marel, R.~P., \& Anderson, J.\ 2010, \apj,
  710, 1063

\end{thebibliography}
\end{document}